\documentclass{sig-alternate-05-2015}

\usepackage[show]{notes-alt}

\usepackage{tabularx}
\usepackage{tabulary}
\usepackage{booktabs}
\usepackage{wrapfig}

\usepackage{float}
\usepackage{url}
\usepackage{tabto}
\usepackage{amsmath}
\usepackage{enumitem}
\usepackage[autostyle]{csquotes}
\usepackage{listings}
\usepackage{subfig}

\usepackage[numbers,square]{natbib}

\newcommand{\hide}[1]{}

\begin{document}

\CopyrightYear{2016} 
\setcopyright{acmcopyright}
\conferenceinfo{SIGIR '16,}{July 17-21, 2016, Pisa, Italy}
\isbn{978-1-4503-4069-4/16/07}\acmPrice{\$15.00}
\doi{http://dx.doi.org/10.1145/2911451.2914724}

\hyphenation{Map-Reduce}
\hyphenation{opti-mi-za-tion}

\title{On the Applicability of Delicious for Temporal Search \\on Web Archives\titlenote{This work is partly funded by the European Research Council under ALEXANDRIA (ERC 339233)}}

\numberofauthors{1}
\author{
\alignauthor
Helge Holzmann, Wolfgang Nejdl, Avishek Anand\\
       \affaddr{L3S Research Center}\\
       \affaddr{Appelstr. 9a}\\
       \affaddr{30167 Hanover, Germany}\\
       \email{\{holzmann, nejdl, anand\}@L3S.de}
}

\maketitle

\begin{abstract}

Web archives are large longitudinal collections that store webpages from the past, which might be missing on the current live Web. Consequently, temporal search over such collections is essential for finding prominent missing webpages and tasks like historical analysis. However, this has been challenging due to the lack of popularity information and proper ground truth to evaluate temporal retrieval models. In this paper we investigate the applicability of external longitudinal resources to identify important and popular websites in the past and analyze the social bookmarking service \textsf{Delicious} for this purpose.

The timestamped bookmarks on \textsf{Delicious} provide explicit cues about popular time periods in the past along with relevant descriptors. These are valuable to identify important documents in the past for a given temporal query. Focusing purely on recall, we analyzed more than 12,000 queries and find that using \textsf{Delicious} yields average recall values from 46\% up to 100\%, when limiting ourselves to the best represented queries in the considered dataset. This constitutes an attractive and low-overhead approach for quick access into Web archives by not dealing with the actual contents.

\end{abstract}

%
%




\section{Introduction}

Most of the information and published content is either online or has been moving to the Web. Consequently, there has been a surge of collection, curation and preservation efforts to archive the live and ephemeral Web. While some institutions, such as the Internet Archive\footnote{\url{http://archive.org}}, attempt to archive the Web comprehensively, more and more national organizations and libraries target the Web under their own top-level domains. The growing number of such endeavors all around the world make Web archives a valuable resource for social scientists, historians, computer scientists as well as for the average user. 

Web archives can provide access to historical information that is absent on the current Web, like previous companies, products, events, entities etc. However, even after a long time of existence, Web archives are still lacking \emph{search capabilities} that make them truly accessible and usable as temporal resources. Available access methods include lookup services, such as the Wayback Machine \footnote{\url{https://github.com/iipc/openwayback}}, or indexing methods for efficient temporal queries~\citep{anand_sigir2011, anand_sigir2012}, but retrieval models to rank versions of webpages are limited to relevance cues from document content~\cite{berberich2010language}. This is primarily due to the inability of the models to determine which page was important at a given instant or interval of time. Furthermore, as there are often multiple versions of a page, it is difficult to identify which version among them is the most relevant and different versions of the same page may be relevant at different points in time. To make matters worse, it is even more difficult to identify the variations of a page that are the most interesting for users in a given time period only by analyzing \textit{internal} properties of a page, like its content, as detailed in~\citep{campos2015survey}.

While determining authority of pages in an archive independent of a query has been attempted~\cite{Nguyen:sigir2015}, popularity cues from external sources have not been considered. By incorporating \textit{external} data, such as explicit temporal information about a website's popularity, this can be simplified and lead to a better retrieval performance. Sources for this can be any websites reporting about other websites, such as social networks, where users post their favorite or most controversial websites at a specific time of interest. Besides the explicit time information, another advantage of searching \textit{external} data instead of websites itself is the more focused descriptions of only relevant pages. Users typically post the essence instead of the often verbose contents found on the websites, including layouts, comments, etc. Finally, this also allows for a much more clean and compact index, a critical factor for ever growing Web archive collections with sizes of hundreds of terabytes.

One of those resources is the social bookmarking service \textsf{Delicious}\footnote{\url{https://del.icio.us}}, where users post popular links and describe them with a concise set of tags as succinct descriptors. These tags carry temporal information that can be exploited for search: While the tag \textit{commuity} is frequently assigned to \textit{facebook.com} today, other communities, such as \textit{myspace.com} were tagged with the term before. The idea to base search on tags has been previously explored, but never in a temporal dimension \citep{BischoffF08, stanford_wsdm08}.

With \textit{Tempas (Temporal archive search)} we have created a prototype to search webpages based on a \textsf{Delicious} dataset over nine years\footnote{\url{http://tempas.L3S.de}}. Tempas links directly into a Web archive to show the versions described by the users in their posts \citep{holzmann_www2016}. We exploit the frequencies of posts as well as mutual information of tags attached to a page as indicator of their relevance to a certain topic. Tempas is a fully functional system to search Web archives in a temporal manner on a relatively small and concise index. However, before we can continue to tune this system to a high precision it is essential to understand what is available in the used dataset and which of the websites that really mattered to users of search engines in the past can be found with it. It is crucial to keep in mind the inherent bias of the dataset used for this task.

In this paper we analyze the applicability of \textsf{Delicious} for searching documents published in the past based on the following research questions: 

\begin{sloppypar}
\begin{enumerate}[noitemsep, nolistsep]
\item[(\textbf{R1})] \textit{What fraction of websites clicked by users on a search engine in the past are included in \textsf{Delicious} as well?}
\item[(\textbf{R2})] \textit{What topics or entities are covered by \textsf{Delicious}?}
\item[(\textbf{R3})] \textit{What are the natural limitations of \textsf{Delicious} as a dataset for temporal Web archive search?}
\item[(\textbf{R4})] \textit{How do posting times on \textsf{Delicious} and query times in search engines relate to each other?}
\end{enumerate}
\end{sloppypar}

\vspace{-1mm}

\section{Analysis}

\begin{figure}
	\centering
	\includegraphics[width=\columnwidth]{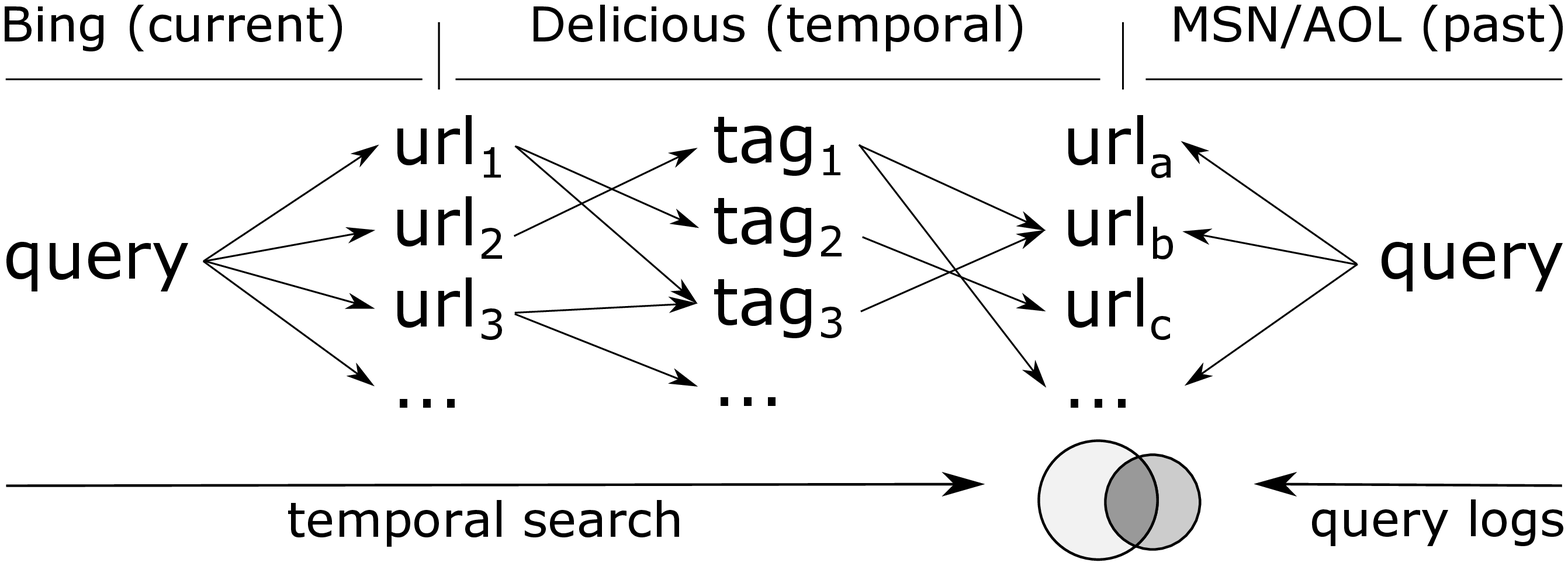}
	\vspace{0mm}
	\caption{The analysis workflow from query-tag mapping using Bing and \textsf{Delicious} to querying \textsf{Delicious} and comparing against MSN/AOL query logs.}
	\vspace{-5mm}
	\label{fig:workflow}
\end{figure}

The \textsf{Delicious} dataset spans nine years from 2003 to 2011 and contains about 340 mio. bookmarks, 119 mio. unique URLs, 15 mio. tags and 2 mio. users \citep{socialbm0311}. Each bookmarked URL is time stamped and tagged with descriptors. The methodology of our analysis is shown in Figure~\ref{fig:workflow}.

We begin with a query workload and identify the associated tags for each query in \textsf{Delicious}. This is done by using Bing search results as a proxy and selecting tags attached to the returned URLs. We then compute the overlap of clicked search results in two query logs, from AOL and MSN, and the URLs tagged with the query or its expanded tags to compute the \emph{recall} of \textsf{Delicious} for temporal search. This is done for overlapping time intervals of both datasets, query logs and \textsf{Delicious}, which are May 2006 for MSN and March to May 2006 for AOL.

As an example scenario consider the clothing brand \texttt{American Apparel}. Using the approach as described below, we map this query to the tags \texttt{americanapparel} and \texttt{apparel} as well as \texttt{t-shirts}, which appears to be used quite synonymously on \textsf{Delicious}. In the next step, we retrieve all URLs that were tagged with any of these tags during a given time, here May 2006. This results in 227 URLs. As a ground truth we compare against query logs from MSN and AOL from the same time. These contain two and four URLs that users clicked on for the query \texttt{american apparel}: \texttt{americanapparel.net} and \texttt{americanapparelstore.com} on MSN as well as \texttt{allonlinecoupons.com} and \texttt{usawear.org} additionally on AOL. As only the first two are contained in the set of URLs from \textsf{Delicious}, it reaches a full recall of 100\% (1.0) with respect to MSN and 50\% (0.5) w.r.t. AOL. Analyzing the precision remains for future work as it will require an appropriate retrieval model to rank the relevant links up to the top. A high recall, however, is a crucial prerequisite to make the system usable in practice.

\subsection{Query-Tag Mapping}
\label{sec:entity_tag_mapping}

Our query workload comprises article titles from the German Wikipedia, which we consider as entities in the following. In principle, the approach is applicable to any type of queries, but the mapping of entity names to tags is more straightforward than arbitrary multi-keyword queries, which would usually refer to a set of tags instead of single representatives. The used Wikipedia collection consists of 1.8 million articles from the main namespace without disambiguation and list pages. The focus on German aligns with the Web archive available to us and also narrows down the vast number of articles on Wikipedia. The titles were issued as queries to Bing between August 7 and 13, 2015 and we stored the top 100 results for each query.

To map the queries to tags that best represent the corresponding entity, we collected all tags from \textsf{Delicious} that were attached to the retrieved URLs from Bing and used by at least 10 users. From these we kept only those tags that were used by at least 10\% of the users who posted one of the URLs for that query. In addition, we selected a reference tag $w_\text{ref}$ tag that exactly matches the query after down-casing and removing special characters. E.g., $w_\text{ref}(\text{Barack Obama}) = \texttt{barackobama}$. For the remaining tags we computed an adaption of \textit{IDF} (inverse document frequency) based on the total number of considered queries and relative to the reference tag for normalization:
\[\text{idf}(w) = \frac{\text{log}(|\text{queries}|)}{|\{q \in \text{queries}\ |\ w \in \text{tags}(q)\}|}\]
\[\text{rel.idf}(w) = \frac{\text{idf}(w)}{\text{idf}(w_\text{ref})}\]
This number indicates the generality of the tag, i.e., how specific it is to the assigned query. Additionally, we computed what we call \textit{exclusiveness}. A high exclusiveness indicates that the tag does not often co-occur with the reference tag, which would be uncommon if both tags represent the same entity:
\[\text{excl}(w) = 1 - \frac{\text{\#posts}(w, w_\text{ref})}{\text{min}(\text{\#posts}(w),\ \text{\#posts}(w_\text{ref}))}\]
$\text{\#posts}(w_1, w_2, ...)$ defines the number of posts on \textsf{Delicious} with all specified tags $w_1, w_2, ...$ as unique pairs of user and posted URL to filter spammers with a large number of posts of the same URL.

To combine the numbers we computed the average score of each tag $w$ as $0.5 \cdot (\text{rel.idf}(w) + \text{excl}(w))$ for every query with reference tag $w_\text{ref}$. The resulting score indicates how specific a tag is to its considered query and at the same time how complementary or exchangeable it is to the reference tag, which we consider a representation of the corresponding entity. Experiments have shown that a threshold of 0.7 reliably identifies true representative tags for a query. Thus, we took all tags with a score equal or greater as well as $w_\text{ref}$. 

Examples are shown in Table~\ref{tab:entity_listing}. Although this simple approach is not accurate in all cases, it is good enough for an experiment like ours. While it worked well for entities like \texttt{Wikipedia}, it sometimes yields too general tags, like \texttt{popular} in case of \texttt{Craigslist}. For other entities, some tags seem too generic at the first glance, but appear to be used almost synonymously on \textsf{Delicious}, such as the format \texttt{flv} for \texttt{YouTube}. However, the quality is acceptable and a better mapping would have only resulted in even higher recall values as we will show in the following.

\subsection{Querying}

\begin{figure}
	\centering
	\vspace{-5mm}
	\includegraphics[width=\columnwidth]{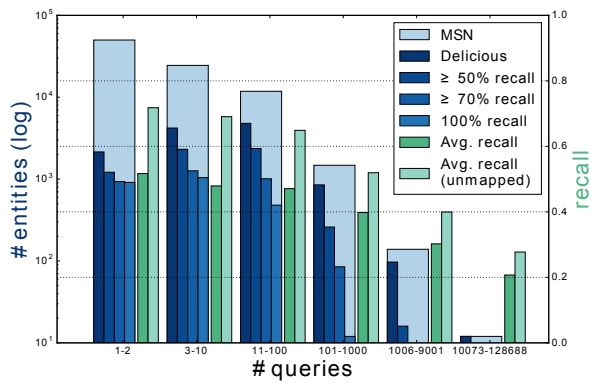}
	\vspace{-3mm}
	\caption{Comparison of the numbers of entities in the MSN query log vs. Declicious with different recalls and the average recall considering entity-tag mapping or not, partitioned by the popularity of entities according to number of queries in the logs.}
	\vspace{-5mm}
	\label{fig:MSN_popularity_recall}
\end{figure}

To assess the recall of the considered \textsf{Delicious} dataset, we \textit{queried} it for the identified tags and at the same time selected records for the corresponding entity from the MSN and AOL query logs. We matched queries with exact name as extracted from the Wikipedia titles, ignoring case. The clicked URLs for a given query served as ground truth in the experiment and we measured recall by counting how many of those could be also found in \textsf{Delicious}. Each URL annotated with one of the considered tags was selected. We operated under the assumption that tags identified by us were also used back in the past, which we believe is valid, as most tags relate to entity names or variations. In total, 12,106 entities could be successfully mapped to tags as well as corresponding queries in the MSN query logs and 12,170 in the AOL logs.

From Table~\ref{tab:entity_listing}, which shows the top 10 entities with highest frequency in the MSN logs and a recall of greater than 0.5 on \textsf{Delicious}, a certain bias towards rather technical and Internet related entities can be observed (\textbf{R2}). A bias to a certain, less popular type of entities could also be affirmed by counting numbers and recall values for entities of different popularities as shown in Figure~\ref{fig:MSN_popularity_recall}: Even though all of the most popular entities could be mapped to tags in \textsf{Delicious}, their recall is relatively low. However, the highest presence in \textsf{Delicious} was not among the least popular entities either, but among those that were queried by up to 100 search sessions in the MSN logs. Also, the highest recall values were reached for the long-tail as well, rather than for the more popular entities (\textbf{R3}).

To ensure the observed recalls are not due to a poor mapping (s. Sec.~\ref{sec:entity_tag_mapping}), we computed the recall values of the same entities also by considering the entire \textsf{Delicious} dataset, not only results retrieved by the mapped tags (\textit{unmapped}). This suggests the upper bound, which could potentially be reached given a better entity-tag mapping. However, as presented in the figure, these recall values are only slightly higher than the ones achieved with our mapping.

\begin{figure}[b]
	\centering
	\vspace{-5mm}
	\includegraphics[width=\columnwidth]{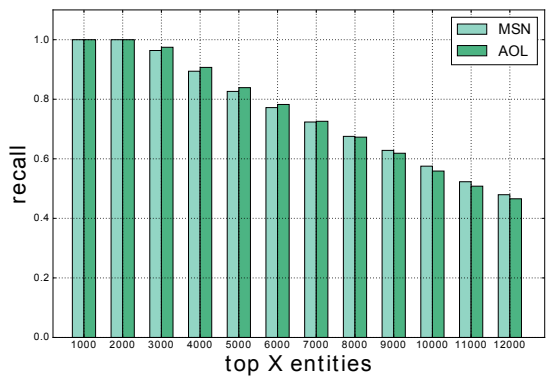}
	\vspace{-3mm}
	\caption{Top X entities according to their \textsf{Delicious} recall with respect to MSN and AOL query logs.}
	\label{fig:Combined_topX_overall}
\end{figure}

Overall, the average recall values of the entire experiment resulted in 48\% w.r.t. MSN and 46\% w.r.t. AOL. Although this is already a good result, it gets even better by looking at those entities, which are strongly represented on \textsf{Delicious} and reached the highest recalls in the experiment. As presented in Figure~\ref{fig:Combined_topX_overall}, for the top 2000 entities we achieve a full recall w.r.t. both query logs. The top 6000, which is about 50\%, still exhibit a recall of almost 0.8 with even slightly higher values w.r.t. AOL (\textbf{R1}).

This shows, about half of the URLs that users clicked in a search engine can be found by using \textsf{Delicious} as an external data source to search Web archives. Even more intriguing, if used with the queries that are best represented by the dataset, we can even reach much higher values up to a recall of 100\%.

\begin{table}[t!]
  \newcolumntype{C}{>{\centering\arraybackslash}X}
  \newcolumntype{L}{>{\raggedright\arraybackslash}X}
  \newcolumntype{R}{>{\raggedleft\arraybackslash}X}
  \small
  \centering
  \caption{Top 10 entities according to MSN query logs from May 2006 with a recall value of greater than 0.5.}
\begin{tabularx}{\columnwidth}{lCrr}
    \toprule
    \textbf{Entity}&\textbf{Tags}&\textbf{\#Q}&\textbf{Recall}\\
    \midrule
ESPN&\texttt{espn}&7492&0.60\\
Gmail&\texttt{gmail}&5285&0.92\\
Craigslist&\texttt{classifieds, popular, imported, craigslist}&4943&0.80\\
Wikipedia&\texttt{wiki, encyclopedia, wikipedia}&3063&0.60\\
Imdb&\texttt{cinema, imdb}&2626&0.92\\
AIM&\texttt{aim}&2180&0.57\\
YouTube&\texttt{flv, converter, youtube}&1965&0.75\\
Sudoku&\texttt{sudoku}&1739&0.75\\
PayPal&\texttt{paypal}&1710&0.60\\
    \bottomrule
  \end{tabularx}
  \vspace{-3mm}
  \label{tab:entity_listing}
\end{table}

\begin{figure*}[ht!]
	\captionsetup[subfigure]{oneside,margin={0cm,0cm}}

	\vspace{-5mm}
	
	\hspace{-2mm}
	\subfloat[all entities (MSN)]{
		\label{fig:MSN_all_12before_12after}
		\includegraphics[width=0.36\textwidth]{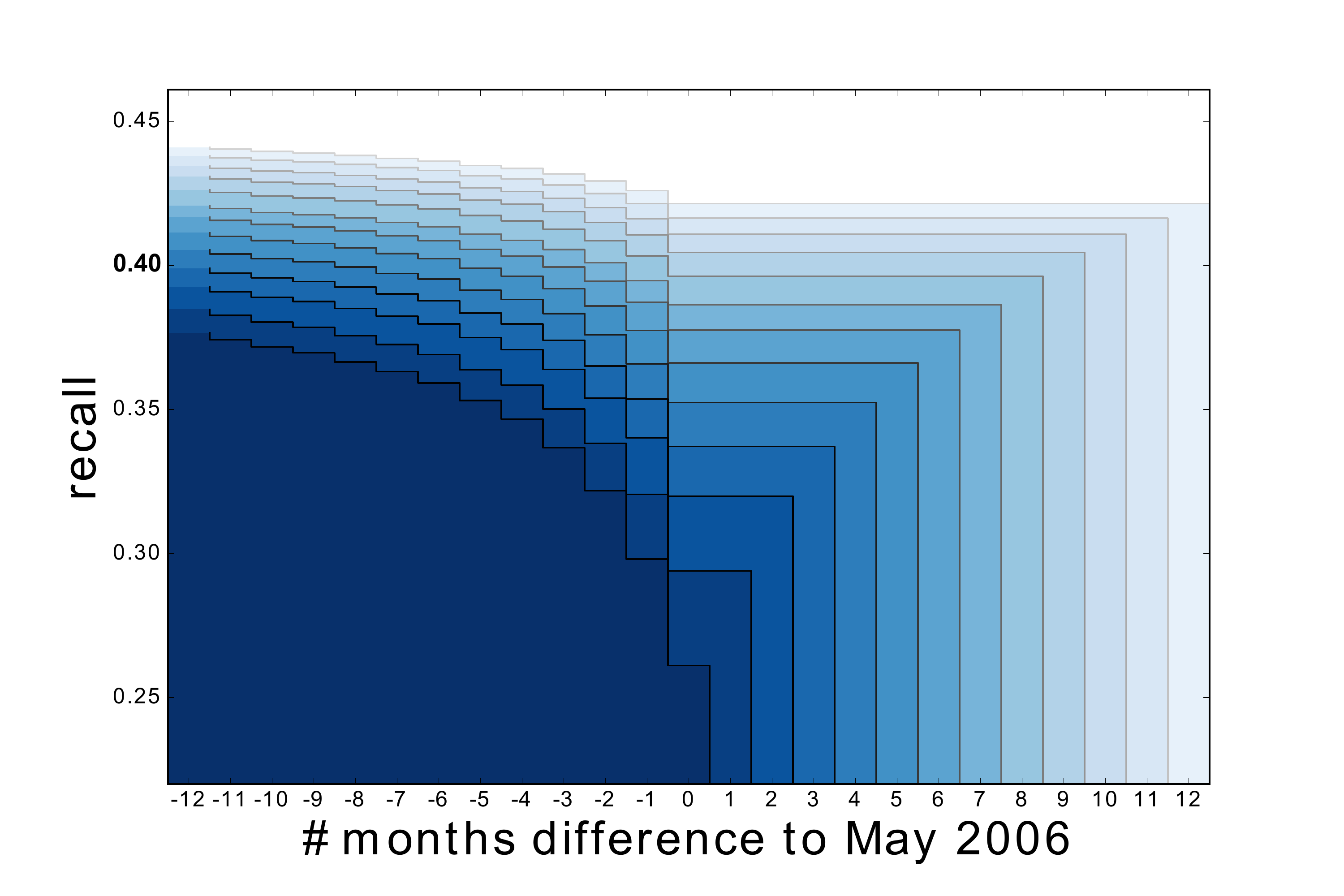}
	}
	~\hspace{-10mm}~
	\subfloat[all entities (AOL)]{
		\label{fig:AOL_all_12before_12after}
		\includegraphics[width=0.36\textwidth]{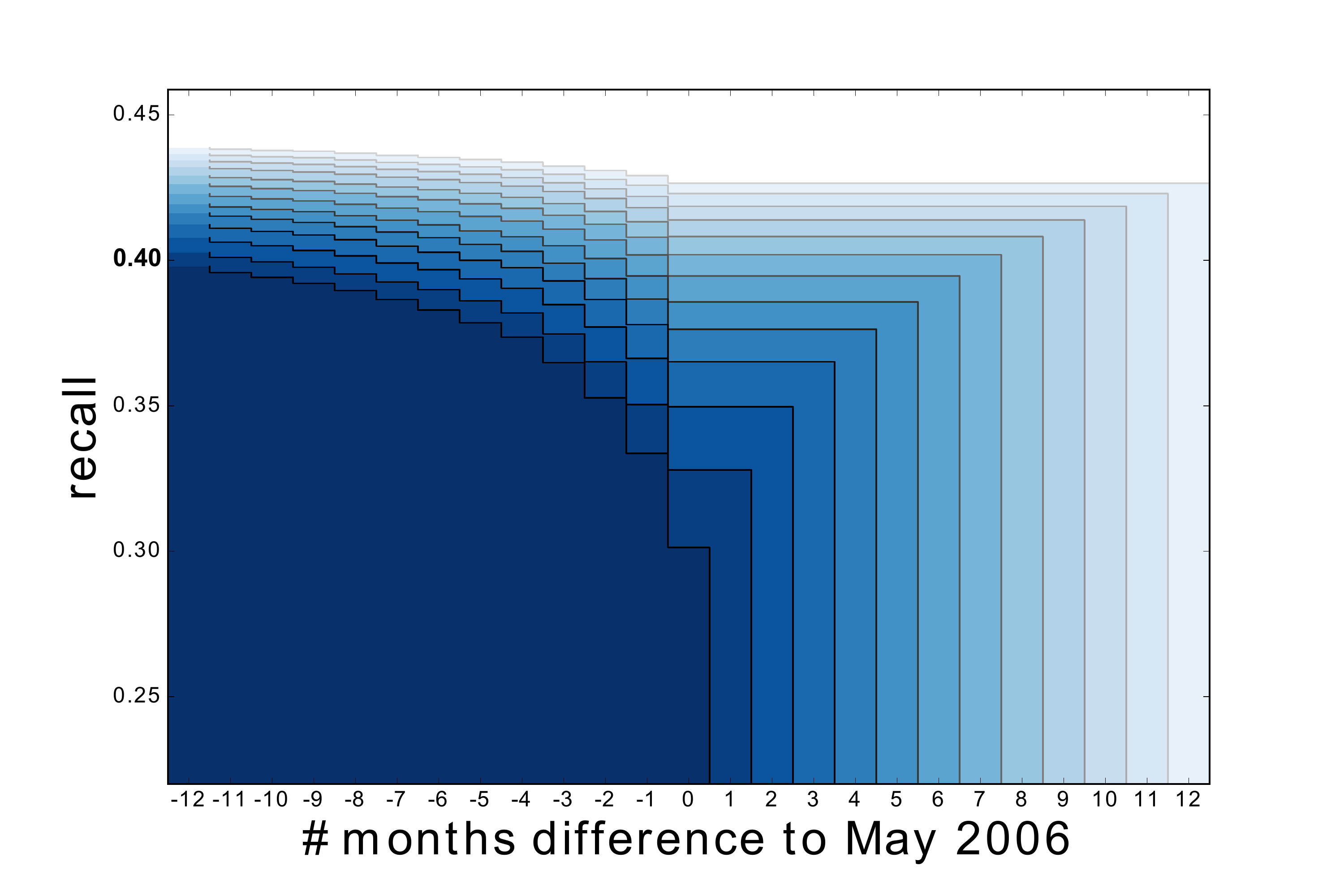}
	}
	~\hspace{-10mm}~
	\subfloat[top 2000 entities (AOL)]{
		\label{fig:AOL_top2000_12before_12after}
		\includegraphics[width=0.36\textwidth]{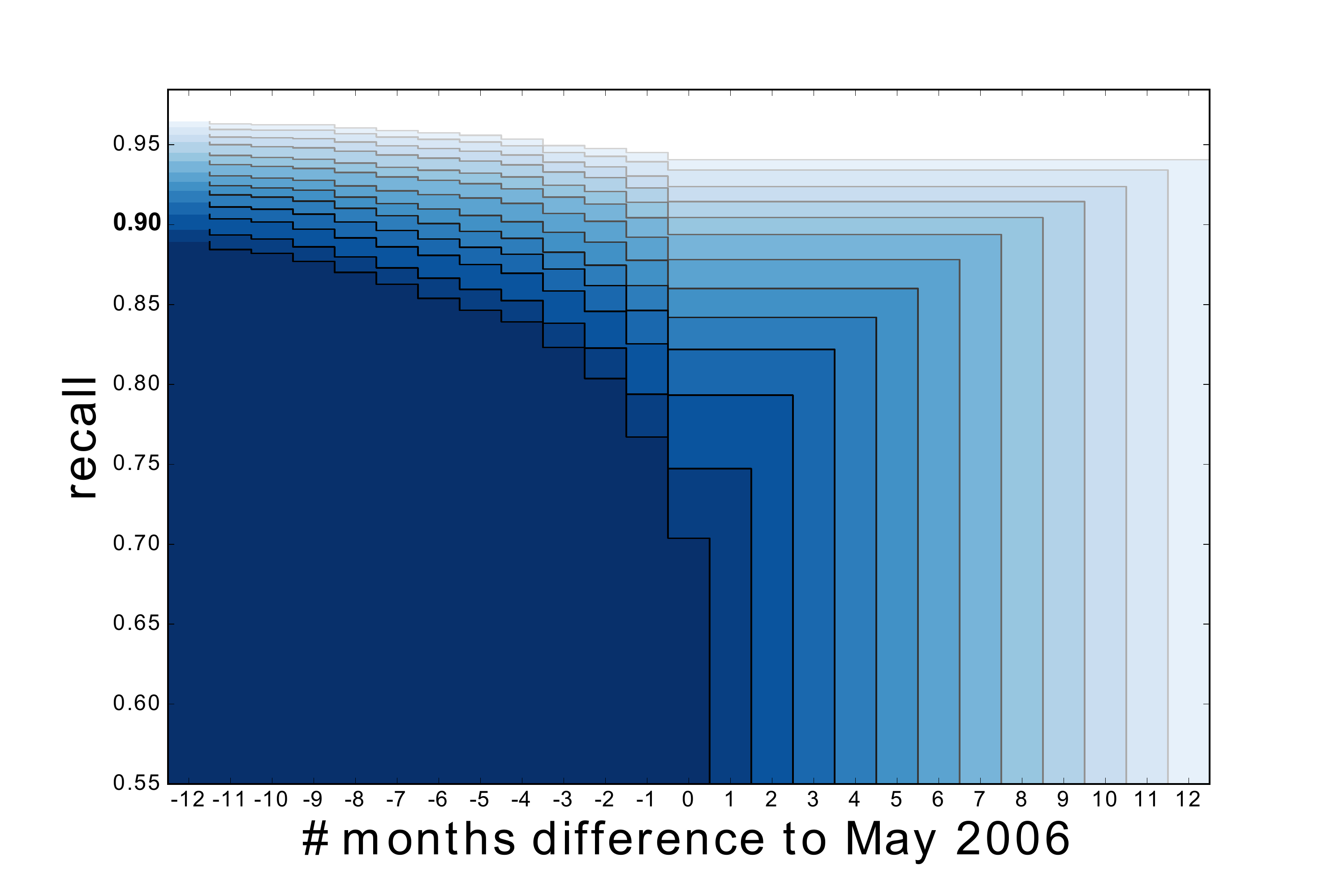}
	}
	
	\vspace{5mm}
	
	\caption{Recall w.r.t. MSN query logs from May 2006 (0) and AOL query logs from March 2006 (-2) to May 2006 (0).}
	
	\vspace{-5mm}
	
	\label{fig:temporal_recall}
\end{figure*}

\section{Temporal Search}

So far, none of the presented analyses has taken into account the temporal aspect, although our ultimate goal was to evaluate the applicability of \textsf{Delicious} as external data source for temporal Web archive search. Thus we wanted to know, given the fairly good recalls achieved before, are the URLs spread across the entire dataset or focused around the query times of the available logs. Figure~\ref{fig:temporal_recall} presents these temporal recall results. To get comparable values, we queried \textsf{Delicious} with the same approach as before just only around the time spans of the logs, i.e., May 2006 for MSN (0) and March (-2) to May 2006 (0) for AOL. The x-axis in the figures denotes the time difference in months up to one year in the future and the past.

Notable is that in all plots, up to a difference of around three months, the recall grows faster by relaxing the search interval to the past. However, from the third month on the recall gain is higher by expanding the search interval to the future. This suggests posts on \textsf{Delicious} are lagging slightly behind the queries in the considered search engines. Overall the results are a slightly higher w.r.t. the AOL query logs, which, however, may be due to the fact that it spans three months instead of one, covered by MSN. Hence, in May \textsf{Delicious} is already two months ahead of AOL. This corresponds to the observation that the recall w.r.t. MSN for three months from 0 to 2 approximately matches the recall w.r.t. AOL in month 0. Overall, the recall results come very close to what we observed in Figure~\ref{fig:Combined_topX_overall}. While we achieved a full recall for the top 2,000 entities from the entire dataset, already a search span of two years, 12 month in the future and 12 in the past, is sufficient to retrieve 95\% w.r.t. MSN and even more than 95\% w.r.t. AOL, as presented in Figure~\ref{fig:AOL_top2000_12before_12after}. Even more intriguing, we only lose about 10\% of recall by searching as little as two months in the past and three months in the future (\textbf{R4}). However, in favor of \textsf{Delicious} was the fact that the platform was relatively popular during the analyzed time. Thus, the results might be lower today and call for alternative datasets.

\section{Conclusion and Outlook}

In this work we introduced the idea of using external data sources with explicit temporal information for Web archive search. We investigated the applicability of the social bookmarking service \textsf{Delicious} as a resource for this task and have shown that it indeed serves as a suitable dataset for a certain kind of entity queries. It is crucial to keep in mind the bias of any dataset used with the presented approach, as obviously only queries supported by the dataset can lead to satisfying results. It turns out, the best represented entities among the most popular ones on \textsf{Delicious} are from the technical or Internet domain. Nevertheless, even overall we reached a promising recall of 46\% to 48\% w.r.t. the analyzed query logs from MSN and AOL. By focusing on the top queries, which exhibit the best recall results, a full recall of 100\% could be retained up to the top 2000 of all of the about 12,100 analyzed entities. The top 50\%, approx. 6000 entities, still result in a recall of almost 80\% on average.

As a second analysis of the paper, we investigated temporal recall values with respect to the time spans of the query logs around May 2006. It has shown that most of the websites of interest w.r.t. the query logs also appeared on \textsf{Delicious} around one year before and after the query times. Already five months around the query time result only in a little loss of recall. This indicates that \textsf{Delicious} is applicable for temporal search as it appears to resemble query logs of the past. The next step in this endeavor will be to create an appropriate retrieval model that ranks the URLs according to their relevance as derived by the query logs. Eventually, our goal is to incorporate different sources, such as different social networks as well as external websites from the Web archive itself, that cover the missing entities over time for a more complete Web archive search.

\vspace{-0.5mm}


\renewcommand*{\bibfont}{\raggedright}

\end{document}